\documentclass{article}

\usepackage[
  paperheight=8.5in,
  paperwidth=5.5in,
  left=10mm,
  right=10mm,
  top=20mm,
  bottom=20mm]{geometry}
\usepackage[utf8]{inputenc}

\usepackage{graphicx}
\usepackage{wrapfig}
\usepackage[bottom]{footmisc}
\usepackage{listings}
\usepackage{pifont}
\usepackage{enumitem}

\usepackage{wrapfig}
\usepackage{ragged2e}
\usepackage{comment}
\usepackage[dvipsnames]{xcolor}

\usepackage{array}
\usepackage{multirow}
\usepackage{booktabs}
\usepackage{hhline}
\definecolor{palegreen}{rgb}{0.6,0.98,0.6}

\usepackage{amsmath}
\usepackage{amssymb}
\usepackage{multicol}
\usepackage{lipsum}
\usepackage{hyphenat}
\PassOptionsToPackage{hyphens}{url}
\usepackage{url}

\usepackage{rotating}

\usepackage[T1]{fontenc}
\usepackage{textcomp}
\usepackage{listings}
\usepackage{setspace}

\newcommand*{\affmark}[1][*]{\textsuperscript{#1}}
\newcommand*{\email}[1]{\small{\texttt{#1}}}

\renewcommand{\footnoterule}{%
  \kern -3pt
  \hrule width \textwidth height 0.5pt
  \kern 2pt
}

\date{}

\usepackage{titlesec}
\titleformat*{\section}{\large\bfseries}
\titleformat*{\subsection}{\normalsize\bfseries}
\titleformat*{\subsubsection}{\normalsize\bfseries}

\title{
\textbf{Designing a Security System Administration Course for Cybersecurity with a Companion Project}\footnote{\protect Copyright \copyright 2023 by the Consortium for Computing Sciences in Colleges.
Permission to copy without fee all or part of this material is granted provided
that the copies are not made or distributed for direct commercial advantage,
the CCSC copyright notice and the title of the publication and its date appear,
and notice is given that copying is by permission of the Consortium for
Computing Sciences in Colleges.  To copy otherwise, or to republish, requires
a fee and/or specific permission.
}
}

\author{
\textit{Fei Zuo, Junghwan Rhee, Myungah Park, Gang Qian}\\
Department of Computer Science\\
University of Central Oklahoma, Edmond, OK 73034\\
\email{\{fzuo, jrhee2, mpark5, gqian\}@uco.edu}\\
}

\begin{document}
\maketitle

\begin{abstract}

In the past few years, an incident response-oriented cybersecurity program has been constructed at University of Central Oklahoma. As a core course in the newly-established curricula, Secure System Administration focuses on the essential knowledge and skill set for system administration. To enrich students with hands-on experience, we also develop a companion coursework project, named \textsc{PowerGrader}. In this paper, we present the course structure as well as the companion project design. Additionally, we survey the pertinent criterion and curriculum requirements from the widely recognized accreditation units. By this means, we demonstrate the importance of a secure system administration course within the context of cybersecurity education.

\end{abstract}

\section{Introduction}

In recent years, we have witnessed an immense shortage of cybersecurity professionals and practitioners across the nation. To fill the gap between the low supply and high demand of the cybersecurity workforce, 
an increasing number of colleges have launched initiatives to establish new cybersecurity programs or courses in their computer science departments.

In 2021, we initiated a cybersecurity degree program and the certificates for both undergraduate and graduate students at our institution~\cite{rhee2023develop}. It is worth noting that our program features \emph{incident response}, which is an essential task to address security incidents and secure infrastructure. This special theme requires practical strength such as certifiable deep knowledge, hands-on skills, and research-involved cybersecurity activities. To this end, we proposed an array of cybersecurity courses to cover the essential knowledge and skill set for the major tasks of incident response.

Within our cybersecurity curriculum, the Secure System Administration (SSA) course plays a significant role. It aims to introduce the core knowledge and skill set for operating and administering systems securely. In particular, students learn and practice system administration techniques to operate a system with script languages in a command-line interface. In this process, the important concepts, components, and terminologies used in policy, regulation, and risk management for secure system administration are also delivered.

Considering that cybersecurity is a practice-oriented discipline, we always emphasize that students should not only understand how everything works theoretically but also be able to apply the related knowledge to solve real-world problems. For this purpose, we have developed a companion coursework project named \textsc{PowerGrader}, which is a \textit{PowerShell}-based system for automatic management and assessment of programming assignments. We anticipate that students could polish their skills in using a script language by developing applications in a concrete scenario. They are also expected to deeply experience how applications of automation and configuration management work through this practical project. 

\section{Course Construction}\label{syllabus}

We develop this course with the objective that upon successful completion of this course, students will gain a solid understanding of essential concepts, components, and terminologies involved in system administration policies, controls, and risk management. They will also be able to proficiently administrate computer systems via command-line interfaces and other productivity tools. Also, 
the capability of designing and implementing the automation of computer operational tasks using script languages is a designated learning outcome. 

\subsection{Course Structure}

Students attending the course are juniors in the CS program as well as
students in the software engineering program. They should have finished and passed the Programming I and Programming II courses or equivalent. Table~\ref{tab01} shows the course structure, which is divided into four topics. All of them are commonly suggested in a syllabus concerning security system administration geared towards those aforementioned objectives.

\begin{table}[t]\footnotesize
\setlength{\abovecaptionskip}{0cm}
\setlength{\belowcaptionskip}{0.4cm}
\caption{Course Structure and Knowledge Units (KUs)}

\centering
\begin{tabular}{|c|l|l|}
\hline
\textbf{Topics}                                        & \multicolumn{1}{c|}{\textbf{Knowledge Units (KUs)}}   &         \multicolumn{1}{c|}{\textbf{Student Activities}}                                                                                                            \\ \hline
\begin{tabular}[c]{@{}c@{}}\textbf{T1}\\ (3 weeks)\end{tabular} & \begin{tabular}[c]{@{}l@{}}• Administration basics\\ • Terminals and CLI\\ • Windows essentials\\ • UNIX/Linux essentials\end{tabular}    & \begin{tabular}[c]{@{}l@{}} Project: Phase 1 \\ Quiz 1  \end{tabular}    \\ \hline
\begin{tabular}[c]{@{}c@{}}\textbf{T2}\\ (3 weeks)\end{tabular} & \begin{tabular}[c]{@{}l@{}}• Regular expression\\ • Editors\\ • Software management \\ • Other system tools\end{tabular}  &      \begin{tabular}[c]{@{}l@{}} Project: Phase 2   \\  Quiz 2  \end{tabular}     \\ \hline
\begin{tabular}[c]{@{}c@{}}\textbf{T3}\\ (5 weeks)\end{tabular} & \begin{tabular}[c]{@{}l@{}}• Script Languages - Bash\\ • Script Languages - PowerShell\end{tabular}  &   \begin{tabular}[c]{@{}l@{}} Project: Phase 3   \\   Quiz 3     \end{tabular}        \\ \hline
\begin{tabular}[c]{@{}c@{}}\textbf{T4}\\ (3 weeks)\end{tabular} & \begin{tabular}[c]{@{}l@{}}• Policies and regulations\\ • Risk analysis and management\\ • Security controls and frameworks\\ • System certifications\end{tabular} & \begin{tabular}[c]{@{}l@{}} Project: Report   \\  Final exam \end{tabular} \\ \hline
\end{tabular}\label{tab01}

\end{table}

\vspace{6pt}
\noindent\textbf{T1: Command-line interface and essential commands}
\vspace{3pt}

System administration is the field of work in which someone manages one or more systems including software, hardware, servers or workstations. Its main goal is to ensure systems are running efficiently and effectively. It is not surprising that command-line interface (CLI) has become a crucial skill for system administrators. Its programmable characteristic provides great convenience for scheduling automation. Therefore, demonstrating proficiency in the usage of CLI is of necessity  for professional management. 

\vspace{6pt}
\noindent\textbf{T2: System tools and productivity applications}
\vspace{3pt}

Mastering good tools are a prerequisite to the success of any jobs. For instance, a regular expression (or shortened as regex) is an immensely powerful tool that can be used to improve productivity in routine tasks. Besides this, other system tools such as Advanced Package Tool (APT), Windows management instrumentation (WMI), and editors (e.g., Emacs or Vim) are covered. 

\vspace{6pt}
\noindent\textbf{T3: Scripting languages and automation}
\vspace{3pt}

Scripting languages, e.g. \textit{PowerShell} and \textit{Bash},  are commonly utilized in system administration tasks due to the convenience provided by them for manipulating and automating operating system facilities.  
As an open-sourced and cross-platform system administration and configuration framework from Microsoft, \textit{PowerShell} is widely used to automate routine and repetitive tasks. For example, \textit{PowerShell} is leveraged to automate configuration management in computer networking laboratories~\cite{palmer2019automating}.
Moreover, current cyber-criminals or hackers usually adopt \textit{PowerShell} as a component of their attack tool-chain. The number of penetration tools that use \textit{PowerShell} increased accordingly at high speed in recent years~\cite{hendler2018detecting}. That is why \textit{PowerShell} has been integrated into many cybersecurity-related courses~\cite{liu2020web}.

\vspace{6pt}
\noindent\textbf{T4: Policies, controls, and risk management}
\vspace{3pt}

In an organization, avoiding risks is often difficult due to many reasons such as human fallibility and uncontrollable external factors. Sometimes it is even impossible. How to assess and manage risks is thus an important capability for a qualified system administrator. In this course, essential concepts, components, and terminologies used in system administration policies, controls, and risk management are introduced. 

\subsection{Comparison with Other Similar Courses}

Admittedly, system administration courses are broadly provided by sibling universities. Some are particularly designed for IT programs~\cite{mohamed2022}, which do not extensively cover security-related topics. Others only skim the surface of automation based on script languages or mainly focus on Bash~\cite{yue2022secure}, although they are included in a cybersecurity program. By contrast, our course design differs in three aspects: the hands-on experience involved companion project, the concentration on \textit{PowerShell}-based administration automation, and the 
industry-leading, certificate-oriented course structure.

\section{Companion Project}\label{sys_design}

The prototype of \textsc{PowerGrader} was inspired by our practical needs in teaching entry-level programming courses~\cite{zuo2023power}. We systematically integrate it into the SSA course as a companion project, so that students can gain a solid understanding of how to make effective use of scripts for automatic system administration. In this section, we present more details about \textsc{PowerGrader}.

\subsection{Background and Motivation}

Automatic code assessment solutions are a category of widely adopted auxiliary teaching applications. They have been developed to evaluate students’ programming assignments for correctness, efficiency, and adherence to the best practices and coding standards. At the very beginning, we initiated our customized code assessment tool \textsc{PowerGrader} in response to the actual demands during our teaching practice. Later, we realized that the development of \textsc{PowerGrader} was suitable to be adapted as a companion project for the SSA course. That was not only because learning how to evaluate the code would be especially beneficial for computer science students, but also because through this project, they could experience firsthand the significant role of scripting in system administration automation.

We observe that a majority of the existing methods are implemented based on black-box testing, such as CodeAssessor\cite{vander2013improving} 
and MOCSIDE\cite{barlow2021mocside}. This sub-category of methods are less flexible and
unsuitable for introductory-level programming courses, where students are usually required to practise a certain language syntax or programming paradigm, so that a plausibly correct output cannot be simply regarded as the indicator of an acceptable solution. 

The ``leap year'' question can be used as a case study to illustrate this concern. In this assignment, students are expected to implement a program to determine whether a certain year is a leap or common year. As a classical question, this exercise aims at asking students to practice \textit{nested branch statements}. In reality, however, we noticed some students included complicated logical expressions with boolean operators to subconsciously bypass the original requirement, as shown in Figure~\ref{fig_c1}. It was obvious that code assessment based on black-box testing would underperform when handling such a case.

\begin{figure}[!ht]
\centering
\setlength{\abovecaptionskip}{0.1cm}
\begin{lstlisting}[
language=C++,
basicstyle=\ttfamily\footnotesize, 
breaklines=true, 
keywordstyle=\bfseries\color{NavyBlue},
stringstyle=\bfseries\color{PineGreen!90!black}, 
columns=flexible,
frame=single,
xleftmargin=.25in,
xrightmargin=.25in
]
if (year%4 != 0 || (year%4 == 0 && year%100 == 0 && year%400 != 0)) {
    cout << "Common year" << endl;
}
if ((year%4 == 0 && year%100 !=0 ) || 
    (year%4 == 0 && year%100 == 0 && year%400 == 0)) {
    cout << "Leap year" << endl;
}
\end{lstlisting}
\caption{Solving the ``leap year'' problem without \textit{nested branch statements}.}
\label{fig_c1}
\end{figure}

On the other hand, approaches based on static analysis such as abstract syntax tree or symbolic execution would be more suitable for such intricate tasks, e.g., vulnerabilities discovery and code quality assessment~\cite{keuning2017code, luo2021westworld}. However, these methods 
would usually introduce high a false-positives rate that preclude using them for grading purposes~\cite{jin2020automatic}. Thus, developing an auto-grader for entry-level programming courses based on their own characteristics appeared to be a more appropriate approach.

\subsection{Project Breakdown}

The core module of \textsc{PowerGrader} is a hybrid code assessor consisting of a black-box tester and a lexical analyzer. Apart from that, the proposed system can periodically collect the submissions from a file server, and automatically manage them. When the assignments are analyzed, assessment reports and compiling information including error messages are recorded into a log. Based on these functionalities, we decompose the development of \textsc{PowerGrader} into three phases, as shown in Table~\ref{tab01}. 

\vspace{6pt}
\noindent\textbf{Phase 1: Assignment collection and pre-processing}
\vspace{3pt}

The assignment collection module is deployed based on a file server where students can submit their work via FTP.
A time stamp is automatically attached to each submission. Meanwhile, user credential and access permission are defined accordingly as a security mechanism. In particular, we assume every submission is archived using a ZIP file, and require the file name be in the format as ``\textit{FirstName\_LastName\_AssignmentNumber.zip}''. Consequently, by parsing the name and meta data of such a ZIP file, we can initialize an assessment report including necessary information such as student name, submission time, etc. After decompressing every ZIP file, the source code will be compiled by invoking a third-party compiler. Any information like error or warning messages during this procedure will be logged. 

All of these system administration tasks are required to be conducted by leveraging CLI-based commands. Hence, this phase is to align with the KUs covered by T1, as shown in Table~\ref{tab01}.

\vspace{6pt}
\noindent\textbf{Phase 2: Regex based lexical analyzer}
\vspace{3pt}

The lexical analyzer is implemented in a lightweight manner, where the given regular expressions parse the source code to determine whether the student’s implementation can fulfill the desired constraints. For example, the regex pattern shown in Figure~\ref{fig_c2} can be used to verify \textit{nested branch statements}. Considering that \textsc{PowerGrader} is specially
designed to handle code submitted as coursework in a
programming class, this lightweight approach is not only feasible but also flexible.

\begin{figure}[!ht]
\centering
\setlength{\abovecaptionskip}{0.1cm}
\begin{lstlisting}[
language={[LaTeX]TeX},
basicstyle=\ttfamily\footnotesize, 
breaklines=true, 
keywordstyle=\color{NavyBlue}\bfseries,
morekeywords={if,else},
%otherkeywords={\\},
%emph={\s}, 
%emphstyle=\color{blue},
texcsstyle=*\color{Rhodamine}\bfseries,
moretexcs=s,
columns=flexible,
frame=single,
xleftmargin=.25in,
xrightmargin=.25in
]
if\s*\([\s\S]*\)\s*\{[\s\S]*
    if\s*\([\s\S]*\)\s*\{[\s\S]*\}\s*
    else\s*\{[\s\S]*\}\s*\}\s*
else\s*\{[\s\S]*\}
\end{lstlisting}
\caption{A regex pattern for \textit{nested branch statements}.}
\label{fig_c2}
\end{figure}

Upon finishing the lexical analyzer, students can sufficiently polish their skills of utilizing productivity tools like regular expressions and editors. As a result, this phase covers the KUs related to T2 in Table~\ref{tab01}.

\vspace{6pt}
\noindent\textbf{Phase 3: Black-box test and automation}
\vspace{3pt}

In the black-box test, expected outputs are literally compared with the counterparts from running a student's code against a test set. In addition to this, all aforementioned functionalities will be integrated into the final product by making effective use of script languages. According to evaluation results obtained from the hybrid code assessment, a report will be generated as a reference for each submission. 

\begin{table}[t]\footnotesize
\setlength{\abovecaptionskip}{0cm}
\setlength{\belowcaptionskip}{0.4cm}
\caption{Agreement to the Course Evaluation Questions}

\centering
\begin{tabular}{|c|c|c|}
\hline
\textbf{Course evaluation question}                                                                                   & \multicolumn{1}{c|}{\textbf{Average score}}  & \multicolumn{1}{c|}{\textbf{Standard error}} \\ \hline
\begin{tabular}[c]{@{}c@{}}This course challenged me to \\ think from different angles\end{tabular}                    & 4.50/5.00    &   0.183                                               \\ \hline
\begin{tabular}[c]{@{}c@{}}I acquired multiple essential \\ system administration skills \end{tabular} & 4.81/5.00 &  0.136                                                  \\ \hline
\begin{tabular}[c]{@{}c@{}}The project is well designed\\ and practiced my skills\end{tabular}                         & 4.75/5.00   &  0.112                                                \\ \hline
\begin{tabular}[c]{@{}c@{}}The techniques delivered by this \\ course are very useful in practice\end{tabular}       & 4.63/5.00   &   0.155                                               \\ \hline
\end{tabular}\label{tab02}

\end{table}

To achieve automation, script language programming is highlighted in this phase, which covers the KUs related to T3 in Table~\ref{tab01}.

\vspace{6pt}
\noindent\textbf{Project report}
\vspace{3pt}

On completion of all the phases above, a closing report will be required to conclude the whole project, where the newly developed application will be considered as a case study. Students apply the knowledge covered by T4 as shown in Table~\ref{tab01} to analyze this coding assignment management and assessment system. Through this exercise, they can sharpen their understanding of security policies, controls, and risk management-related concepts in practice.

\subsection{Evaluation}

Engaging in this hands-on project can make students deeply experience how an automatic system administration and configuration application works in a concrete scenario. Plus, a majority of knowledge units required in the course are covered by this comprehensive project. 

\begin{figure}[h]
\centering
\includegraphics[width=0.6\textwidth]{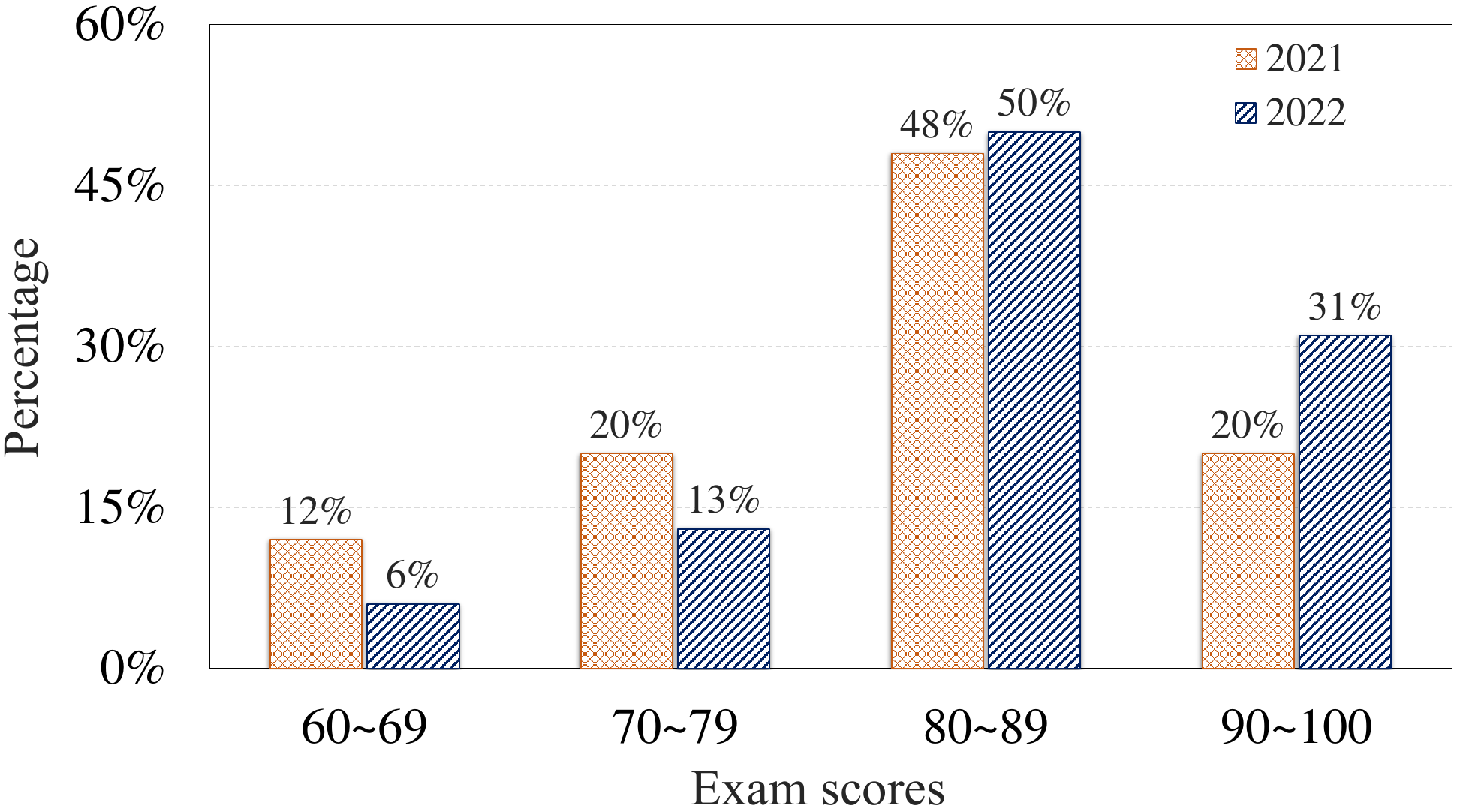}
\caption{Comparison of exam score distributions in different years}
\label{figure:bar_chart}
\end{figure}

To evaluate the proposed course including the companion project, we collected student responses to the course evaluation at the end of 2022. Based on 16 valid responses, Table~\ref{tab02} summarizes their degree of agreement to the questions, which demonstrates a positive feedback. 

In addition, we compared students' performances according to their exam scores in two consecutive academic years. The distribution of scores is shown in Figure~\ref{figure:bar_chart}. The average score was promoted from 83 in the year 2021 to 88 in 2022. Hence, a notable performance improvement has been observed after the course project's implementation in 2022.

\section{Survey on Reputable Accreditation Criteria}\label{relate}

\subsection{NCAE-C Program}

The National Centers of Academic Excellence in Cybersecurity  program, managed by National Security Agency (NSA), accredits universities based on their ability to meet rigorous academic criteria and offer top-notch cybersecurity education and training to students~\cite{cae}. The program has established standards for cybersecurity curriculum and academic excellence towards two specialization tracks, which are Cyber Defense (CAE-CD) and Cyber Operations (CAE-CO). It is noteworthy that the CAE-CD designates five technical core knowledge units (KUs) and another five non-technical core KUs as the curriculum requirements. The proposed SSA course covers two non-technical core KUs, i.e., ``security risk analysis'', and ``policy, legal, ethics, and compliance'', in addition to one technical core KU, i.e., ``basic scripting and programming''.

\subsection{ABET Accreditation for Cybersecurity}

ABET is a well recognized organization that accredits university programs in natural science, computing, engineering and engineering technology. In particular, the Computing Accreditation Commission of ABET has developed the accreditation criteria for cybersecurity programs~\cite{abet2023}. The criteria of ABET explicitly state that curriculum requirements do not prescribe specific courses. Instead, the program needs to involve eight fundamental topics in total, and specifies what course(s) each topic is covered. Our SSA course provides coverage on multiple cybersecurity topics as required by ABET, including data security, human security, organizational security, and societal security. These required topics overlap with T4 of our course structure. Our investigation shows that other ABET accredited cybersecurity programs also consider the SSA course as a necessity~\cite{yue2022secure}.

\subsection{CompTIA Certification}
To meet the needs of the industry, we also
refer to criteria and credentials expected in the job market as the curricular guidance when creating our cybersecurity program. In particular, our department became a partner of CompTIA~\cite{comptia}, which is a well-known trade association that issues professional certifications for the cybersecurity and IT industry. Our curriculum is closely aligned with CompTIA certificates. Especially, the proposed SSA course covers the majority of the topics related to the CompTIA Linux\texttt{+} certificate and some topics for the CompTIA Security\texttt{+} certificate. 

\section{Conclusion}\label{conclude}

As a core course in many cybersecurity-related curricula, secure system administration focuses on not only the essential concepts used in
security policies, controls, and risk management, but also the necessary techniques that are used for automating system administration and configuration. Our SSA course has all these suggested topics combined in the development of the \textsc{PowerGrader} course project to enrich students with hands-on experience in automated system administration and further improve their learning outcomes. 
This paper also surveys the widely recognized accreditation criteria to demonstrate the significance of the SSA course in a cybersecurity program. More importantly, this work shares our experience and insights into teaching SSA and provides an example of its curriculum construction.

\medskip


\bibliographystyle{plain}
\bibliography{sample}

\end{document}